# Dynamic exciton funneling by local strain control in a monolayer semiconductor

Hyowon Moon[1,*], Gabriele Grosso[2], Chitraleema Chakraborty[1], Cheng Peng[1], Takashi Taniguchi[3], Kenji Watanabe[3], and Dirk Englund[1,†]

[1]Department of Electrical Engineering and Computer Science, Massachusetts Institute of Technology, Cambridge, MA, USA
[2]Photonics Initiative, Advanced Science Research Center, City University of New York, New York, NY, USA
[3]National Institute for Materials Science, Tsukuba, Ibaraki, Japan

Corresponding authors: *hwmoon@mit.edu ; †englund@mit.edu

**The ability to control excitons in semiconductors underlies numerous proposed applications, from excitonic circuits for computing and communications[1,2] to polariton condensates[3] to energy transport in photovoltaics[4]. 2D semiconductors are particularly promising for room-temperature applications due to their large exciton binding energy[5,6]. Their enormous stretchability gives rise to a strain-engineerable bandgap that has been used to induce static exciton flux in predetermined structures[7–10]. However, dynamic control of exciton flux represents an outstanding challenge. Here, we introduce a method to tune the bandgap of suspended 2D semiconductors by applying a local strain gradient with a nanoscale tip. This strain allows us to locally and reversibly shift the exciton energy and to steer the exciton flux over micron-scale distances, as observed by wide-field imaging and time-resolved photoluminescence spectroscopy. We anticipate that the ability to strongly and dynamically modulate the bandgap of a semiconductor at the nanoscale not only marks an important experimental tool but will also open a broad range of new applications from information processing to energy conversion.**

A recurring challenge in many opto-electronic systems is the need to interconvert information or energy between the optical or electronic domains. A promising approach is the use of excitons -- bound states of electrons and holes -- as information carriers intermediate between electronic and optical modalities, or as carriers in energy conversion systems[2,11]. Since charge-neutral excitons are weakly affected by external in-plane electric fields, previous studies have sought to control the long-lived interlayer exciton[12] (with electron and hole confined in different quantum wells) through the quantum confined Stark effect (QCSE) in a bulk material AlAs/GaAs at low temperature (~100 K)[1,2]. Two-dimensional (2D) transition metal dichalcogenides (TMDs) provide a platform for room-temperature excitonic and polaritonic devices, thanks to the large exciton binding energy in such materials[13,14]. Recent work demonstrated gate-controlled propagation and polarization control of interlayer excitons in 2D TMD heterostructures[15,16].

The unusually high elasticity of atomically thin materials offers an alternative: controlling both intralayer and interlayer excitonic flux by strain-based bandgap engineering[17–19]. Moreover, the relatively long lifetime of intralayer excitons in TMDs at room temperature (1-4 ns)[20] suggests that strain-induced bandgap gradients can funnel them up to several microns[4], comparable to recent demonstrations with interlayer excitons[15]. Strain-based exciton funneling was recently observed in fixed geometries including nanostructures and wrinkles[7–10]. However, the ability to reversibly control exciton transport is a key step to illuminate range of new applications, from broad-band solar cells to efficient and compact excitonic devices. Here, we address this problem through in-situ application of a local strain gradient to modulate bandgap. By optical spectroscopy at cryogenic and room temperatures, we show that the applied strain allows exciton funneling into single spot in any spatial direction and position without the need of complex heterostructure assembly or electrical gates.

The working principle of the experiment is illustrated in Fig. 1a. A cantilever (shaded blue) with a sharp tip (sub-10 nm radius) induces a strain gradient in a TMD membrane suspended on a transmission electron microscopy (TEM) grid with 900 nm radius holes. We focus here on monolayer $WSe_2$, though the technique can be extended on any 2D materials. The cantilever is controlled by a 3-axis piezoelectric translation stage (see Method). As seen in the simulation of exciton energy shift in Fig. 1b, the tip-induced strain creates a local depression in the exciton potential energy, $\Delta E(r)$, which produces a funneling force $F(r) = -E(r)/dr$, where $r = \sqrt{x^2 + y^2}$.

We first discuss cryogenic (4K) measurements. Figure 2a plots strain-dependent photoluminescent (PL) spectra. The tip is initially at the highest position (piezo voltage 70V), giving maximum strain to $WSe_2$. It is progressively lowered to reduce strain (10V), and then raised again to the original position (70V). The blue-shaded region in Supplementary Fig. 1 shows the calculated displacement and strain range. The symmetry of the emission spectra about the minimum-strain condition (white dashed line) demonstrates the reversibility of the process. The green dashed line at 1.73 eV shows the blueshift (redshift) of the free-exciton peak ($X^0$) as the strain decreases (increases). PL from a localized band (LB) at lower energy, which we attribute to impurity-bound excitons or local strain potentials, emerges at higher strain levels. The four PL spectra in Fig. 2b, acquired at different strain values, show the dramatic increase of LB intensity and energy shifts of the free-exciton peaks. Similar tuning effects are achievable with in-plane motion of the tip after indentation with the TMD monolayer (Supplementary Fig. 2).

Figure 2c plots the effect of the strain on the free-exciton energy (green) and linewidth (red). Tensile biaxial strain causes a linear redshift of the emission energy up to 12 meV, while the linewidth broadens by 70% (from 13 to 22 meV). We attribute this linewidth broadening to the distributed PL energy across the diffraction-limited spot (>500 nm), as explained in the following paragraph. The applied strain also modulates the emission intensity. Figure 2d shows that the total emission intensity, $I_{total}(V)$, increases 47 ± 9.6 % for the highest voltage (brown), in agreement with previous works[8,9]. The fraction of PL intensity above 1.65 eV, $I_{E>1.65\ eV}(V)$ (orange squares) is constant for different strain values, suggesting that the funneled excitons fall into the localized band of lower energy before recombining (Supplementary Fig. 3). The nonlinear energy shift of the LB is stronger than for free excitons and can be used to estimate the strain in the absence of the free exciton peak (Supplementary Fig. 4).

Figure 3 visualizes the strain-induced exciton funneling in the suspended WSe$_2$ membrane. Figure 3a shows the spatial distribution of the free exciton energy shift due to the indentation, $\Delta E(x,y) = E_{50V}(x,y) - E_{0V}(x,y)$. The energy at each position is obtained by rastering the pump laser position at 175 nm increments across the suspending hole. Figure 3b plots the corresponding PL intensity change, $\Delta I(x,y) = I_{50V}(x,y) - I_{0V}(x,y)$. Both energy shift $\Delta E(x,y)$ and PL intensity change $\Delta I(x,y)$ reach their maximum at the tip position, consistent with excitons drifting and then recombining in the region of maximum strain.

We now compare the experimental data with a theoretical strain model. The strain applied on the membrane with the nanoindentation at the center of the circular hole follows the nonlinear equation[21–23] (see Method):

$$F_{tip} = \sigma_0^{2D} \pi \delta + E^{2D} \frac{q^3 \delta^3}{r_{hole}^2}, \quad \varepsilon_{max}^{2D} = \sqrt{\frac{F_{tip}}{4\pi r_{tip} E^{2D}}} \tag{1}$$

where $\delta$ is the maximum deflection of the membrane in the out-of-plane direction, $\sigma$ is the prestress in the membrane, $E^{2D}$ is the two-dimensional Young's modulus, $q$ is a dimensionless constant calculated from Poisson's ratio $v$ as $q = 1/(1.05-0.15v-0.16v^2)$, and $r_{hole}$ and $r_{tip}$ are the radii of the hole and the tip, respectively. This model predicts a maximum strain of $\varepsilon_{max}^{2D} = 3.7\%$ at the tip location (Supplementary Fig. 5a). However, because the diffraction-limited collection spot size is two orders of magnitude larger than the tip diameter, the local spectral shift is washed out and the optical response shows broad distribution of much smaller strain values (Supplementary Fig. 5b). Then, the estimated energy shift (green dashed line) agrees with the measured energy shift profile (black solid line) as shown in Fig. 3c. The linewidth broadening can also be estimated using this distribution. Figure 3d assumes the unstrained free exciton emission at zero energy with an initial linewidth of 13.5 meV (black line). The optically averaged emission profile at the maximum strain (50 V) is plotted as a green solid line, showing the linewidth broadening of 4.5 meV, which match the experimental results displayed in Fig. 2c.

More than an order of magnitude longer exciton lifetime and drift length at room temperature lend support to the interpretation to exciton funneling. Figure 4a shows the reversible PL modulation by controlling the external strain. The emission energy, linewidth, and intensity are plotted in Fig. 4b, showing an energy redshift and intensity increase similar to the cryogenic case. The stark contrast is the linewidth narrowing at larger strain, which has been explained as the reduced exciton-phonon coupling[24]. Time-resolved PL measurement with pulsed laser excitation shows the drift of the exciton density in Fig. 4c. The upper panel shows the profile of the normalized exciton density without external strain. Excitons are dragged away from the excitation spot presumably due to intrinsic strain in the suspended membrane. The time-dependent broadening is attributed to the diffusion process. However, the drift of excitons can be fully controlled when external strain is applied. Bottom panel of Fig. 4c shows that excitons generated at the center funnel into the tip position, highlighted by a white dashed line. In Supplementary Fig. 6, we map the exciton density in the presence of local strain-induced potential, as derived from the diffusion equation:

$$D\frac{\partial^2 n}{\partial x^2} + \mu\frac{\partial}{\partial x}(n\frac{\partial \varphi}{\partial x}) + G - \frac{n}{\tau} = \frac{\partial n}{\partial t}, \qquad (2)$$

where $D$, $\mu$, $n$, and $\tau$ are the diffusion coefficient, mobility, concentration, and lifetime of the excitons; $\varphi$ is the potential energy modulated by nanoscale-tip induced strain; and $G$ is the generation rate of the excitons (see Method). The value of $D$ and $\tau$ are extracted from the upper panel of Fig. 4c as 0.97 cm$^2$/s and 1.02 ns, which are very well match to the previously reported values[25,26]. Figure 4d shows wide-field images of the normalized exciton emission taken without (top) and with (middle) indentation. The emission intensity difference between the two cases is shown in the bottom panel and illustrates the new emission spot created by indentation. The emission intensity at the tip position can be actively modulated with strain control, as demonstrated in Fig. 4e. The vertical profile of the emission intensity by indentation is plotted in Fig. 4f as a black line. It is fitted with two Gaussian functions centered around the excitation (dotted blue line) and at the tip position (dotted red line). The excitation laser spot preserves its circular symmetry even in the case of indentation, confirming that the observed PL signal results from the funneling process rather than scattered light from the tip (Supplementary Fig. 7).

The distance between the excitation spot and tip position has been varied in Fig. 4g. The CCD images of the exciton emission at different distances are fitted into two Gaussian functions: one centered around the excitation position and the other around the tip position. The relative position of the funneled exciton with respect to the excitation spot is displayed as green dots, demonstrating that the deviated emission spot follows the indented position. The same measurement has been done in the orthogonal axis (red), showing that we can actively control the direction and magnitude of exciton funneling with this technique. The dot size represents the emission intensity of the funneled excitons, and exponentially decreases with distance (inset). The characteristic lengths of exciton drift with this strain profile are 361 ± 48 nm and 314 ± 29 nm in each direction, which implies that 5% of the exciton can drift up to 1 μm at room temperature. The full evolution of the exciton emission for the increasing distances is displayed in Supplementary Fig. 8.

In conclusion, we introduced a method for strong, dynamic modulation of the bandgap of a semiconductor at the nanometer scale. This method enables the control of exciton funneling in atomically thin semiconductors through a strain-modulated local bandgap gradient at both cryogenic and room temperatures. Hyperspectral photoluminescence images show the spatially varying exciton energy and the emission intensity as a function of applied strain. Wide-field CCD image and time-resolved PL measurements demonstrate the control of exciton flux direction, spatial position, and the magnitude of the exciton funneling over hundreds of nanometers at room temperature. Extremely long lifetime (10 ns) of super-acid treated TMDs[27] and large diffusion coefficient (14.5 cm$^2$/s) in hBN encapsulated WSe$_2$[28] indicate that exciton funneling over an order of magnitude longer distance should be possible. These results show the possibility to deterministically and reversibly modulate the bandgap of a semiconductor, providing an important research tool and impacting a broad range of applications from ultrafast excitonic circuits[15] to funneling-assisted photovoltaics[4], deterministically tuned quantum emitters[29], to flux control in excitonic-polariton devices[3].

## Methods

**Sample preparation and active strain modulation**

Monolayer $WSe_2$ was mechanically exfoliated on the polymer substrate, and dry-transferred onto the commercial transmission electron microscopy (TEM) grid with 1.8-µm-diameter holes (TEMWindows). The PL map shows large flakes spanning over several holes (Supplementary Fig. 9). The suspended region emits much stronger PL intensity compared to the non-suspended region, as has already been reported[30]. For room temperature measurement, thin hBN layers (around 10-nm thick) were transferred before $WSe_2$ to support the monolayer on the larger hole size (5-µm-diameter) TEM grid (TedPella). A cantilever with a nanoscale-tip was mounted on the scanning and stepping piezo stages for the positioning (Attocube Inc.). Both the samples and piezo stages in the cryostat (Montana Instrument) cooled down to nominal temperature of 4 K. The cantilever applied strain from the bottom, not to block the excitation and emission light. Finally, the nanoscale tip (qp-SCONT, Nanosensors) was characterized by scanning electron microscopy (SEM) in Supplementary Fig. 10. An AFM tip with higher spring constant was used for room temperature measurement (AC160TS-R3, Olympus).

**Photoluminescence measurements**

All the optical measurement was conducted with a home-built confocal microscopy (Supplementary Fig. 11). A long-working-distance objective lens (Mitutoyo, NA = 0.55, WD = 12 mm) enabled the optical access to the sample from outside the cryostat through an optically transparent window. A different objective lens with higher NA (Olympus, NA = 0.8) was used for room temperature measurement to achieve higher resolution. Two-axis galvanometer mirrors scanned the excitation beam from a tunable laser source ($M^2$ SolsTiS) to image the sample. Collected photons were measured through the spectrometer or single-photon counting module. Wide-field images were acquired with a CCD camera (Thorlabs) before the pinhole, which enabled real-time monitoring of the exciton funneling. A pixel distance of 5.2 µm with the magnification of 111 gives a spatial resolution of 47 nm. Time-resolved photoluminescence data were obtained using a time-correlated single-photon counting module (PicoHarp 300), which is connected to an avalanche photodiode. A supercontinuum laser at 600 nm excitation wavelength was used for the pulsed excitation with a repetition rate of 7.8 MHz, pulse duration of 200 fs, and average power of 1.75 µW. A pinhole in the collection path moved with differential 3-axis translation stage (Thorlabs) in the lateral direction to scan the emission intensity from different position while the excitation spot was fixed.

**Modelling strain distribution**

The force applied to the small tip by a thin membrane follows the nonlinear equation:[21–23]

$$F = \sigma_0^{2D} \pi \delta + E^{2D} \frac{q^3 \delta^3}{r^2}$$

where $\delta$ is the maximum deflection of the membrane in the out-of-plane direction, $\sigma$ is the prestress in the membrane, $E^{2D}$ is the two-dimensional Young's modulus, a dimensionless constant q is calculated from Poisson's ratio $v$ as $q = 1/(1.05-0.15v-0.16v^2)$, and r is the radius of the hole. This force F can be measured by a deflection of the soft cantilever with a spring constant of 0.01 N/m. Two reasonable assumptions simplify the calculation: (1) the deflection of the cantilever is much larger than that of 2D material, and (2) the initial stress is small enough to be neglected. The z-axis piezo stage moves at 41 nm/V at 4 K, resulting in the cantilever deflection of 2.05 µm and corresponding loaded force of 20.5 nN at 50V. Using the mechanical characteristics of $WSe_2$ (Young's modulus of 167.3 GPa[31] and Poisson's ratio of 0.19[32]), we get the maximum deflection of the suspended material $\delta$ = 53.0 nm, verifying the first assumption. The maximum stress of 4.37 N/m and maximum biaxial strain of 3.7 % can be achieved by[21–23]

$$\sigma_{max}^{2D} = \sqrt{\frac{F_{max}E^{2D}}{4\pi r_{tip}}}, \quad \varepsilon_{max}^{2D} = \frac{\sigma_{max}^{2D}}{E^{2D}}$$

where $r_{tip}$ is the tip radius and assumed as 10 nm. Loaded stress on the material is much larger than the initial stress in the suspended membrane on the order of 0.01 N/m[21], verifying the second assumption. Finally, the strain at an arbitrary position (x,y) within the hole is calculated by $\varepsilon = \varepsilon_{max}^{2D} r_{tip}/\sqrt{x^2 + y^2}$, showing the extremely confined strain profile near the center due to 1/r dependence[4]. The calculated strain distribution is displayed in Supplementary Fig. 5a with the same amount of indentation at the center of the hole. However, our optical measurements blur the strain distribution due to the large diffraction-limited spot size of the setup ($\lambda/2NA$ = 664 nm for $\lambda$ = 730 nm, NA = 0.55) despite the nanoscale size of the actual strain gradient. Supplementary Fig. 5b shows the optically averaged map of strain distribution which is obtained by a convolution of the original data and the gaussian-approximated beam profile. The averaged biaxial strain of 0.16% in the central region corresponds to the exciton energy shift of 10 meV[33].

**Numerical simulation of the exciton dynamics**

The drift and diffusion of the excitons can be calculated in time-domain by one-dimensional diffusion equation:

$$D\frac{\partial^2 n}{\partial x^2} + \mu\frac{\partial}{\partial x}\left(n\frac{\partial \varphi}{\partial x}\right) + G - \frac{n}{\tau} = \frac{\partial n}{\partial t}$$

where $D$, $\mu$, $n$, and $\tau$ are the diffusion coefficient, mobility, concentration, and lifetime of the excitons, $\varphi$ is the potential energy modulated by nanoscale-tip induced strain, and $G$ is the generation rate of the excitons. The mobility $\mu$ is related to $D$ by the Einstein equation $D/\mu = k_B T$, where $k_B T$ is the thermal energy. The pulsed laser generates excitons at sub-picosecond duration, then the exciton density behavior is numerically simulated by the finite-difference time-domain method (FDTD) with 1 nm and 1 ps resolution. Finally, the calculated density is convoluted by the gaussian-shaped function in time- and spatial-domain to include finite resolution of the optical measurement setup.

## Acknowledgments

This work was supported in part by the Army Research Office (ARO) Multidisciplinary University Research Initiative (MURI) program, grant no. W911NF-18-1-0431, and in part by the National Science Foundation EFRI 2-DARE, award abstract no. 1542863. Growth of hexagonal boron nitride crystals was supported by the Elemental Strategy Initiative conducted by the MEXT, Japan and the CREST (JPMJCR15F3), JST. H.M. acknowledges support by Samsung Scholarship.

## Author Contributions

D.E. and H.M. conceived the experiment. H.M. and G.G. built the optical measurement setup. H.M. and C.C. fabricated the device. H.M. performed the optical measurement and analyzed the data. P.C. contributed on the sample characterization. T.T. and K.W. supplied the hBN. H.M., G.G., C.C., D.E. discussed the result and wrote the manuscript with the input from all authors.

## Competing interests

The authors declare no competing interests.

# Figures

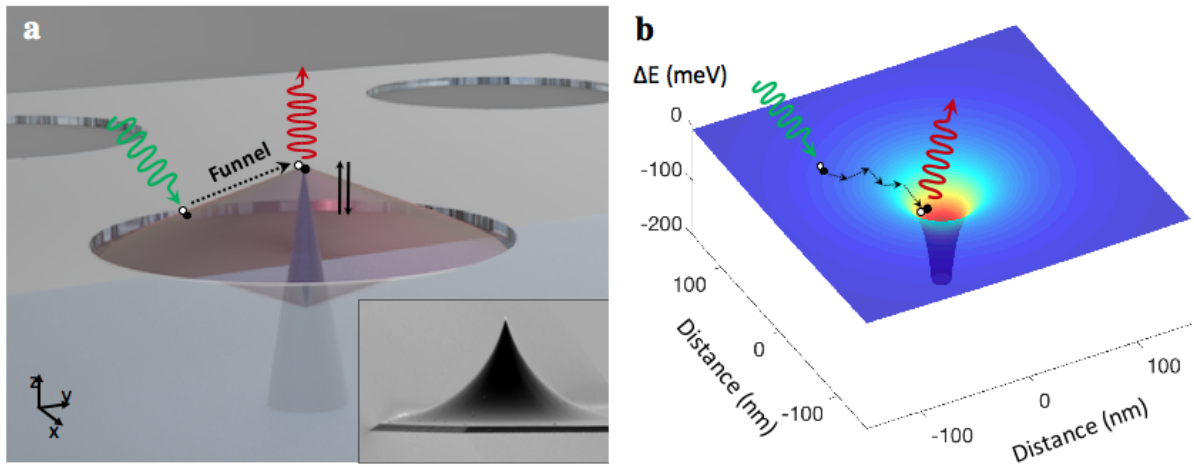

**Fig. 1 | Dynamic local strain tuning for exciton funneling. a,** A schematic shows dynamic exciton flux control in a suspended two-dimensional semiconductor. The piezoelectric translation stage (not shown) moves the blue nanoscale tip to modulate the bandgap of the targeted region. Inset shows SEM image of the real tip. **b,** The optical bandgap of the material decreases as the biaxial strain increases, causing spatially varying energy distribution. The excitons funnel into the highly strained region before recombining.

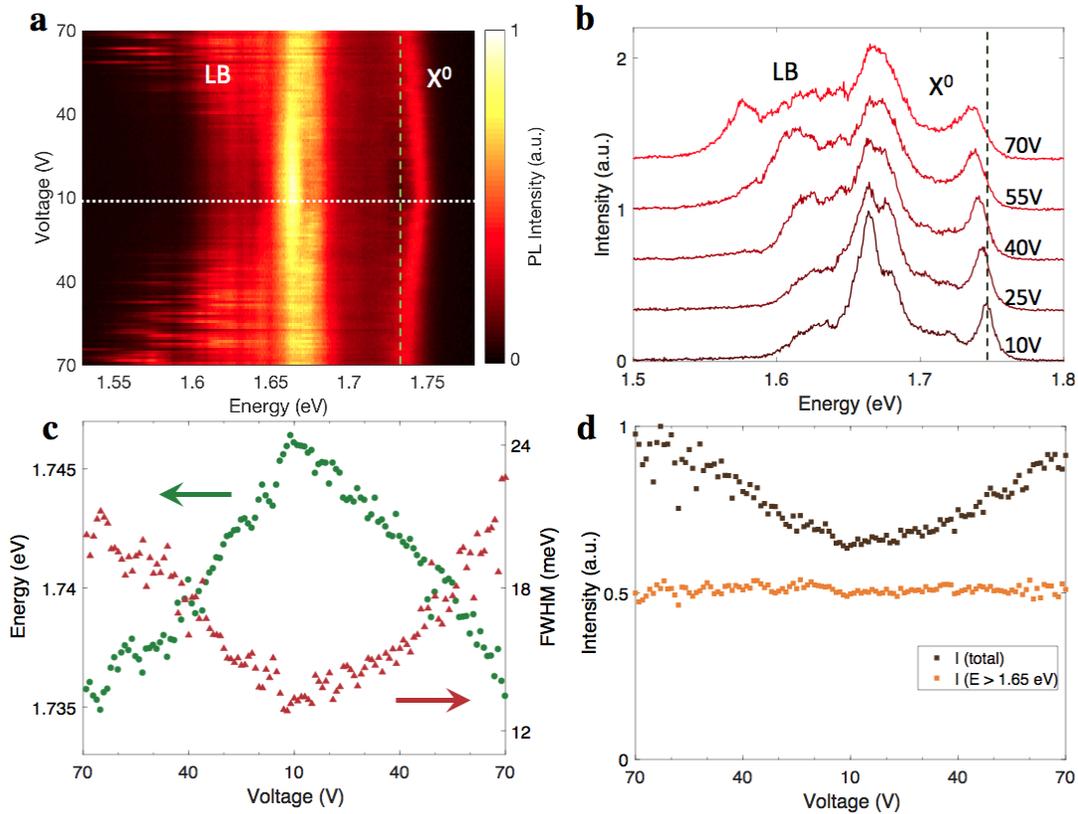

**Fig. 2 | Low temperature strain tuning. a,** Spectral modulation of the free exciton and localized band emissions by a nanoscale-tip induced strain. The voltage of the z-axis piezoelectric stage is decreased from 70V to 10V and returned to 70V. **b,** Spectra with different voltages show redshift in both free-exciton and localized band as the strain increases, in addition to the strong emergence of the lower energy states. **c,** Free exciton energy (green) and linewidth (red) show reversible change with the indentation process. The linewidth broadening is due to the large strain distribution within the diffraction-limited spot. **d,** The strain also modifies the total emission intensity (brown) at the tip while the partial intensity of higher energy above 1.65 eV (orange) remains the same. The increased amount is originated mainly from the lower energy band due to the fast decay process of the funneled excitons.

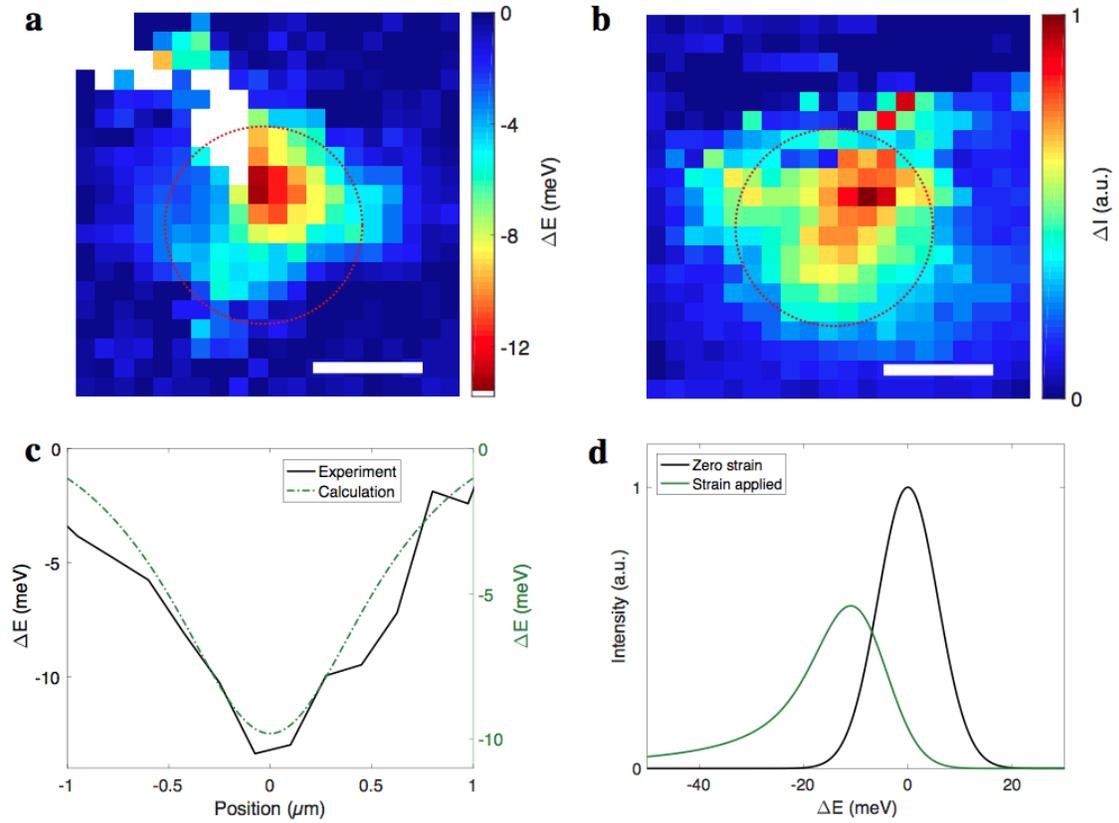

**Fig. 3 | Exciton funneling at low temperature. a,** Spatial distribution of the free exciton energy shift induced by the indentation of 50 V. The free-exciton peak in the white-shaded regions is too weak for reliable fitting and left blank in the colormap. **b,** Similar spatial map that shows the difference of the total emission intensity after the indentation. Intensity also changes gradually in space with the maximum point overlapping with maximum-strain region, demonstrating exciton funneling into the highest strain region. Red circles in **a**,**b** indicate the circumference of the hole, and scale bar is 1 μm. **c,** Linecut of the experimental energy shift distribution (black) and optically-averaged calculated energy shift (green) explains the broadened response well. **d,** Estimated optical response calculated from the strain map convoluted with the gaussian-approximated laser spot size. Linewidth broadening from 13.5 meV to 18 meV explains the experimental result.

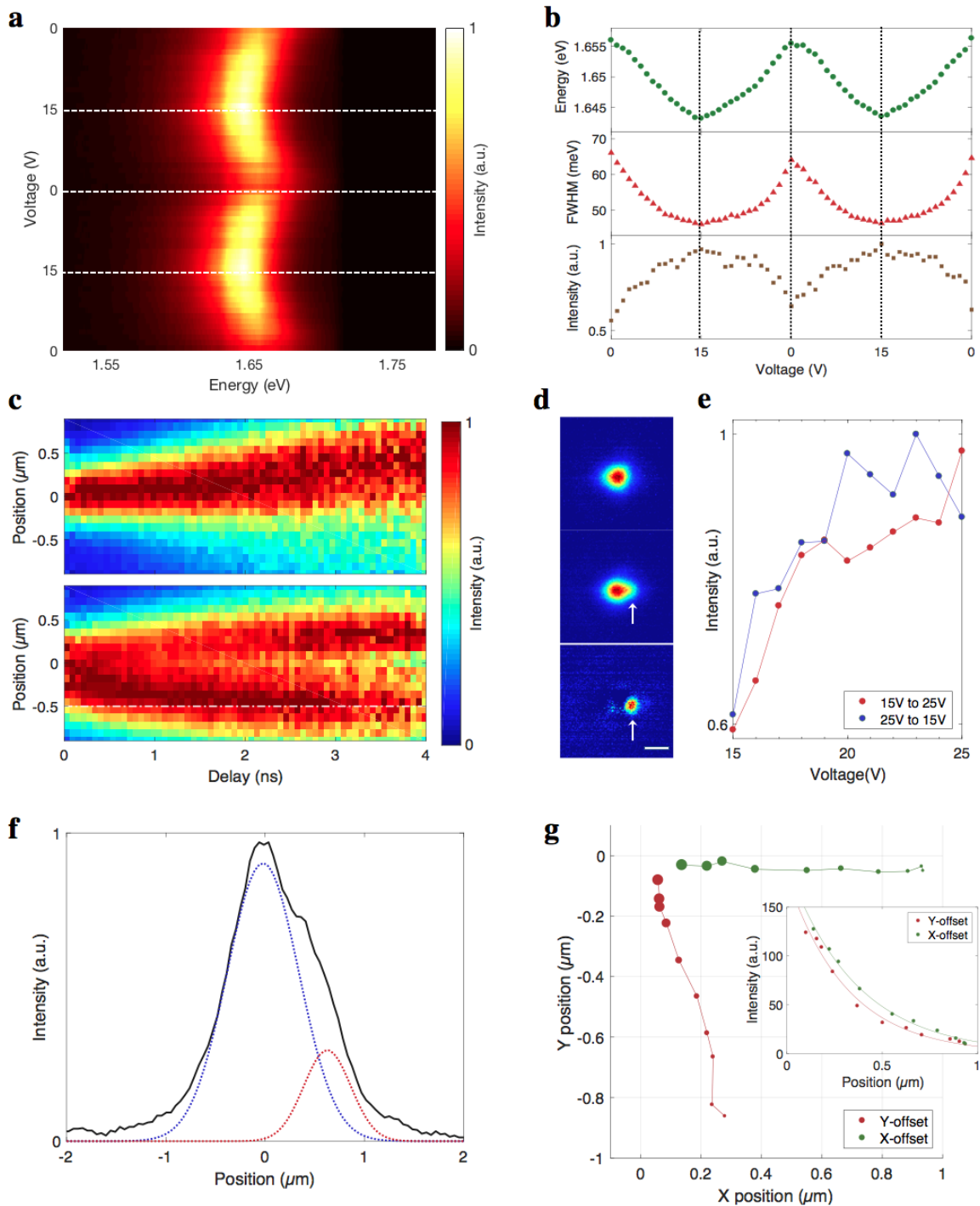

**Fig. 4 | Strain tuning and controllable exciton funneling at room temperature. a,** Reversible spectral modulation of the exciton emission at room temperature. **b,** The exciton energy (green), linewidth (red), and intensity (brown) change with the strain control. Large strain induces the redshift of emission energy, linewidth narrowing, and intensity increase. The linewidth becomes narrow due to the reduced

exciton-phonon interaction. **c,** The lower panel shows the time-resolved exciton density profile with indentation. The excitons funnel into the tip position with indentation (the white dashed line shows the tip position at -0.5 µm). The upper panel shows the control experiment without indentation, showing the initial strain profile funnels excitons into the opposite direction. The emission profile gets broadened due to the diffusion. **d,** Strain modulation of the PL intensity at the tip position 0.5 µm away from the excitation spot. The intensity of the funneled exciton is reversibly controllable. **e,** Wide-field images show normalized PL intensity distribution without (top) and with (middle) indentation, and the difference between the two images (bottom). The tip position is marked as the white arrow, the scale bar is 1 µm. **f**, The black solid line shows the linecut of the middle image of **d**. The dotted lines are Gaussian-fitted functions showing the emission at the excitation spot (blue) and at the funneled spot (red). **g,** The position of the funneled exciton emission relative to the excitation spot, as the excitation spot moves by 100 nm steps in x-direction (green) and in y-direction (red). The size of each dot represents the PL intensity of the funneled excitons, which is plotted and then fitted by a single exponential decay in the inset. A characteristic decay length is longer than 300 nm in both directions.